\documentclass{aa}
\begin{document}
\title{Do X-ray Binary Spectral State Transition Luminosities Vary?} 

\author{Thomas J. Maccarone\inst{}}

   \offprints{T.J. Maccarone; tjm@science.uva.nl}

\institute{Astrophysics Sector, Scuola Internazionale Superiore di Studi Avanzati, via Beirut, n. 2-4, Trieste, Italy, 34014\\ and\\Astronomical Institute ``Anton Pannekoek,'' University of Amsterdam,Kruislaan 403,1098 SJ Amsterdam,The Netherlands}

\date{}

\titlerunning{State Transition Luminosities}
\authorrunning{Maccarone}

\abstract{
We tabulate the luminosities of the soft-to-hard state transitions of
all X-ray binaries for which there exist good X-ray flux measurements
at the time of the transition, good distance estimates, and good mass
estimates for the compact star.  We show that the state transition
luminosities are at about 1-4\% of the Eddington rate, markedly
smaller than those typically quoted in the literature, with a mean
value of 2\%.  Only the black hole candidate GRO~J~1655-40 and the
neutron star systems Aql X-1 and 4U 1728-34 have measured state
transition luminosities inconsistent with this value at the 1$\sigma$
level.  GRO~J~1655-40, in particular, shows a state transition
luminosity below the mean value for the other sources at the $4\sigma$
level.  This result, combined with the known inner disk inclination
angle (the disk is nearly parallel to the line of sight) from
GRO~J~1655-40's relativistic jets suggest that the hard X-ray emitting
region in GRO~J~1655-40 can have a velocity of no more than about
$\beta=0.68$, with a most likely value of about $\beta=0.52$, and a
minimum speed of $\beta=0.45$, assuming that the variations in state
transition luminosities are solely due to relativistic beaming
effects. The variance in the state transition luminosities suggests an
emission region with a velocity of $\sim$0.2$c$.  The results are
discussed in terms of different emission models for the low/hard
state.  We also discuss the implications for measuring the
dimensionless viscosity parameter $\alpha$.  We also find that if its
state transitions occur at typical luminosities, then GX 339-4 is
likely to be at a distance of at least 7.6 kpc, much further than
typically quoted estimates.
}

\maketitle

\section{Introduction}
Early on, it was discovered that the spectral energy distributions of
X-ray binaries showed variations with luminosity.  At low
luminosities, these systems typically show hard X-ray spectra,
dominated by a power-law like component with a photon spectral index
of about 1.8 and a cutoff at a few hundred keV (the low/hard state).
At higher luminosities, they typically show spectra dominated by a
soft quasi-thermal component with a characteristic temperature of
about 1 keV (the high/soft state).  These quasi-thermal spectra are
generally fairly well described by models of a geometrically thin,
optically thick accretion disk (Shakura \& Sunyaev 1973; Novikov \&
Thorne 1973).  The low/hard state spectra are usually described in
terms of thermal Comptonization models in a geometrically thick,
optically thin medium (e.g. Shapiro, Lightman \& Eardley 1976).  The
optically thin region has an electron temperature of $\sim$ 70 keV,
with the high temperature maintained either by magnetic reconnections
(Haardt \& Maraschi 1991; Nayakshin \& Melia 1997; Di Matteo et
al. 1999) or simply by inefficient cooling of the gas at particularly
low accretion rates (e.g. Rees et al. 1982; Narayan \& Yi 1994). Some
recent work has suggested the alternative possibility that the X-ray
emission in at least some of these objects may be dominated by
synchrotron emission from a mildly relativistic jet (Markoff et al.
2001).

Most observed black hole X-ray binaries (as well as a smaller fraction
of the observed neutron star binaries) show spectral state transitions
between the low/hard and high/soft states.  Furthermore, corresponding
radio state transitions are seen as well, with the radio emission
turning on in the X-ray low/hard state and off in the X-ray high/soft
state (Tananbaum et al. 1972; Harmon et al. 1995; Fender et al. 1999).
Finally the best fitting reflection parameters drop substantially as
one goes from the high/soft state to the low/hard state (Zdziarski et
al. 1999), which is usually interpreted as indicative of a ``hole''
developing in the accretion disk (e.g. Poutanen et al. 1997; Esin et
al. 1997), but may also be explained by a mildly relativistic corona
in the low/hard state beaming the hard x-rays away from the accretion
disk (Beloborodov 1999), a jet doing the same thing (Markoff et
al. 2003), or as a result of ionization effects which cause an
artifical correlation between reflection fraction and spectral shape
to appear (Done \& Nayakshin 2001; Ballantyne et al. 2001).
Understanding what happens during the state transitions thus holds
great potential for helping us understand the radiation mechanisms and
the accretion geometry in the individual spectral states (see
e.g. Esin et al. 1997; Poutanen et al. 1997; Merloni 2003).

A systematic study of the luminosities at which these transitions
occur and the possible variations in these luminosities is a
necessary, but generally unavailable piece of the puzzle for
understanding state transitions.  Flux measurements at or near the
state transitions and comparisons of the low/hard state and high/soft
state luminosities have been presented for some individual sources
(see e.g. Zhang et al. 1997; Zdziarski et al. 2002 as well as the
observations discussed below), and the transition luminosity has been
estimated in Eddington units for one source (Sobczak et al. 2000), but
never before has a large sample been collected and analyzed in one
place.  The general lore has held that state transitions occur at
about 8\% of the Eddington luminosity (Esin et al.  1997), and that
the state transition luminosities do not vary much.  Recent work
has shown hysteresis effects in many low mass X-ray binary systems,
where the soft-to-hard state transitions occur at higher luminosities
than the hard-to-soft state transitions (e.g. Miyamoto et al. 1995;
Nowak et al. 2002; Barret \& Olive 2002; Maccarone \& Coppi 2003a).
Furthermore, since some models for the low/hard state come from
regions with bulk relativistic velocities away from the accretion disk
(e.g. Beloborodov 1999; Markoff et al. 2001), while others come from
regions without such motions (e.g. Shapiro et al.  1976; Rees et
al. 1982; Narayan \& Yi 1994), the presence of inclination angle
effects on state transition luminosities may be an important
diagnostic for understanding the accretion geometry of the low/hard
states.  With these aims in mind, we collect from the literature
and/or derive from archival data the masses, distances, and state
transition fluxes for all x-ray binaries where such data is available
and reliable.  We discuss the observations used in Section 2 and
discuss the implications of the mean value, the variance in the values
and the possible effects of inclination angle in Section 3.

\section{Observations}

We have compiled from the literature the data for the sources where
the mass of the compact object, distance to the binary system, and
state transition luminosity have all been measured.  Where the
distances come from optical measurements of the mass donor, we use the
quoted errors.  We have also included several neutron star sources
whose distances have been measured from the luminosities of their type
I bursts.  For these sources, we have set the distance estimate errors
to be 30\% in accordance with the results of Kuulkers et al. (2003).
The distance uncertainties are discussed in greater detail below.  We
also discuss below which sources which are known transients were not
included in our sample and why they have not been included.

For a few additional sources, the data exist except for the
state transition flux.  In these cases, we have estimated the state
transition flux, either from existing fits to the data or by
downloading the appropriate data from HEASARC and fitting it.  The
sources of data for each system are discussed below, and the relevant
parameters are all listed in Table 1.

We have, in general, chosen the data with the best temporal sampling
among data sets capable of measuring the state transition.  In some
cases, this means using a narrower bandpass instrument than the most
broadband instrument which observed the state transition.  This choice
is justified, however, by the fact that the luminosities can change
rather quickly for X-ray transients, but the spectral shapes of
low/hard state black hole accretors are rather constant.  That is to
say, the difference in luminosity caused by observing a source a week
after its state transition is generally larger than the uncertainty
introduced by extrapolating 2-20 keV data in order to measure the
bolometric luminosity.  There is greater variation in the spectral
shapes of the neutron star accretors, but given that {\it RXTE} has
generally provided the best broadband spectroscopy {\it and} the best
temporal sampling, the choice need not be made for the neutron star
sources included in our sample.

We have focused here on the soft-to-hard state transitions.  There are
two major reasons for this.  Firstly, the luminosities of X-ray
binaries in outbursts often fit a fast-rise/exponential decay profile,
so, with the source luminosity changing more slowly during the
decaying than during the rising part of the outburst, errors in
determining the exact time of the state transition will cause smaller
errors in the overall luminosity.  Secondly, hysteresis has been found
to be ubiquitous in the state transition luminosities of X-ray
binaries, with the transition from the hard state to the soft state
occurring at a higher luminosity that the soft-to-hard state
transition.  Since one possible explanation for this hysteresis effect
is that the soft-to-hard state transitions occur in quasi-equilibrium
states, while the hard-to-soft occur in violently unstable states,
there may be instrinsic variations in the hard-to-soft state
transition luminosities that do not occur in the soft-to-hard state
transition luminosities (Maccarone \& Coppi 2003a).  

{\it Nova Muscae 1991.}  The measurement of the X-ray flux is taken
from the {\it Ginga} All-Sky Monitor light curve of Kitamoto et al. (1992),
with the state transition estimated to have occurred on day 135.  The
black hole mass and distance estimates come from Gelino et al. (2001).

{\it XTE J 1550-564.} The measurement of the X-ray flux comes from
Sobczak et al. (2000), with the assumption that the state transition
occurred at the luminosity measured in observation number 205.  The
black hole mass and distance estimate come from Orosz et al. (2002).
The state transition luminosity was estimated by Sobczak et al. (2000)
to occur at about 0.02 $L_{EDD}$, but the estimate was made before the
mass and distance of the black hole had been measured and bolometric
corrections to that luminosity estimate were not made.

{\it GS 2000+251.} Like Nova Mus 91, the measurement of the X-ray flux
is taken from the {\it Ginga} All-Sky Monitor light curve of Kitamoto et
al. (1992), with the state transition estimated to have occurred on
day 135.  The black hole mass and distance estimates come from
Callanan et al. (1996).

{\it Cyg X-1.}  The X-ray flux measurements come from a fit to the
spectral data from the September 2, 1996 data, the first low/hard
state observed by RXTE after the high/soft state seen in the
spring/summer of 1996.  The 3-20 keV spectrum shows a flux of
1.5$\times10^{-8}$ ergs/second/cm$^2$ when a $\Gamma=1.9$
absorbed power law plus Gaussian (to fit the iron emission line) model
is fit.  The mass measurement comes from Herrero et al. (1995),
assuming an inclination angle of 30 degrees and the distance estimate
comes from from Gies \& Bolton (1986).

{\it GRO~J~1655-40.} The measurement of the X-ray flux comes from
Sobczak et al. (1999), with the state transition occurring on August
14, 1997.  The mass estimates come from Greene et al. (2001).  We note
that the high proper motions seen in the relativistic jets of
GRO~J~1655-40 (Hjellming and Rupen 1995), place a firm upper limit on
the distance of 3.5 kpc (Fender 2003).  The distance determinations
for this sources thus have two constraints - one from the optical
measurements, which suggest that $d=3.8\pm0.7$ kpc (Greene et al.
2001), and the other that $d<3.5$ kpc (Fender 2003).  We therefore
combine these two pieces of information and find that $d=3.3\pm0.2$
kpc.

{\it LMC X-3}.  This source was long thought to be steadily in the
high/soft state until the discovery of occasional, brief state
transitions by Wilms et al. (2001).  Luminosities are not quoted for
the state transitions, but the dates during which the source was in
the hard state are identified by Wilms et al. (2001).  The hard state
observation closest to the soft-to-hard transition was taken on May
29, 1998 (RXTE ObsID 30087-02-07-00).  We have downloaded the data for
this ObsID and fit an absorbed power law model to the PCA and HEXTE
data from 3-200 keV, using the standard HEASARC screening criteria.
The best fitting model gives a power law index of $1.8\pm0.1$ and a
flux of $5.0\times10^{-11}$ ergs/sec/cm$^{2}$.  The distance is
assumed to be the standard distance to the Large Magellenic Cloud, 51
kpc (Keller \& Wood 2002).  As a persistent, high mass system,
ellipsoidal light curve variations have not been measured in LMC X-3,
so the system's inclination angle is not well-constrained.  The black
hole's mass estimates come from Cowley et al. (1983) and Paczynski
(1983).

{\it Aql X-1.}  This system is an accreting neutron star, and its mass
is assumed to be the standard 1.4 $M_\odot$ values for neutron stars.
The flux at the state transition comes from Maccarone \& Coppi
(2003a).  The distance of this system is a debated point; the
estimate in the paper presenting the first optical spectrum is 2.5 kpc
(Chevalier et al. 1999) while other work has suggested a distance of
4-6.5 kpc (Rutledge et al. 2001).  We have adopted the smaller value
here as it is based entirely on optical measurements, but also present
alternative calculations for the higher value.

{\it 4U 1608-52.}  The distance to 4U 1608-52 is estimated to be 3.6
kpc on the basis of radius expansion type I X-ray bursts (Nakamura et
al. 1989).  As a neutron star, the mass of the central object is
assumed to be 1.4 solar masses.  The flux is taken from the
well-sampled RXTE PCA/HEXTE data set in the November 2001 outburst.
The source was first observed in a hard spectral state early on 20
November 2001 (ObsID 60052-02-06-00).  The first observed hard state
flux is 2.7$\times10^{-9}$ ergs/sec/cm$^2$ (2-20 keV); after
correcting for neutral hydrogen absorption and making the bolometric
spectral correction (the spectrum has a photon index $\Gamma=1.72$,
and a cutoff energy of 50.6 keV), we find that the bolometric flux is
about 5.6$\times10^{-9}$ ergs/sec/cm$^{2}$.  At the quoted distance of
3.6 kpc, the luminosity is 8.1$\times10^{36}$ ergs/sec, which
corresponds to 4.2\% of the Eddington luminosity for a 1.4 $M_\odot$
neutron star.

{\it 4U 1728-34.}  The distance to this source is estimated to be
$4.3\pm0.5$ kpc on the basis of neutron star atmosphere modeling of a
sample of Type I bursts (Foster et al. 1986 - FRF).  The state
transition with the best temporal sampling is found to have occurred
on 22 February 1996.  The flux at this transition is 2-200 keV flux at
this transition is found to be $4.6\times10^{-9}$ ergs/sec/cm$^2$,
with an optical depth of 0.75 and a temperature of 36 keV for the
Comptonizing region.  The bolometric luminosity is then
9.6$\times10^{36}$ ergs/sec, corresponding to 5\% of the Eddington
luminosity, assuming that the model for the distance estimate is
correct.  Additional confidence in the distance estimate was ascribed
to its consistency with the distance of a putative globular cluster
associated with 4U 1728-34 (Grindlay \& Hertz 1981), whose existence
has since been refuted (van Paradijs \& Isaacman 1989).  Given the
lack of this confirmation of the distance estimate and the fact that
FRF did not consider the effects of changing the chemical composition
of the accreted gas, we have increased the uncertainty in the distance
estimate to this source to the 30\% found by Kuulkers et al. (2003) to
be the rough systematic error.

\subsection{Sources not included}

There are several soft X-ray transients with suspected black hole
primaries which are not included in this work.  The reasons vary -
typically there is either no mass estimate, no distance estimate, or
the X-ray data at the time of the state transitions lack either the
necessary spectral or spatial resolution.  We note that this is much
more likely to be the case for neutron star systems than black hole
systems; the state transitions and the changes in luminosity are much
more rapid for neutron stars than for black holes, perhaps because the
timescale for luminosity change scales with the mass of the accreting
object; therefore, while sampling a few times a week may be sufficient
to measure the state transitions of black hole systems, it will
generally not be sufficient for the neutron star systems.  This is
compounded by the fact that the sampling of black hole systems has
generally been better at the times of state transitions than the
sampling of the neutron star systems, presumably because the black
hole systems have been much brighter in the all-sky monitors and hence
have attracted more attention.  There are also several neutron star
sources not included in the sample for various reasons - generally
because a transition to a full low/hard state was not seen with
sufficient temporal sampling, or because of extreme uncertainties in
the source distance.  Below we discuss the observations for all
sources listed as atoll sources in the most recent ``van Paradijs
catalog'' (Liu et al. 2001) that have not been included in the
analysis as described above.  Only the atoll sources are included
because the Z sources are thought to all be accreting steadily at
luminosities well above the soft/hard transition level and the
unclassified neutron star sources are unclassified for the simple
reason that they have not shown spectral state phenemenology.

\subsubsection{Neutron stars not included}

{\it 4U 1705-44.}  The transition flux for this source has been
estimated to be between $7\times10^{36}$ and $2.1\times10^{37}$
ergs/sec (Barret \& Olive 2003), with the distance assumed to be 7.4
kpc (as estimated from non-radius expansion bursts by Haberl \&
Titarchuk 1995).  However, it seems from the spectral fits to the data
presented in this paper, that a full low/hard state is not reached;
the optical depth in the Comptonization model (using the COMPTT model
in XSPEC; Hua \& Titarchuk 1995) never drops below 5.5 and the
temperature never rises above 14.1 keV (compare, for example with Aql
X-1 where the Comptonization model fits show a drop in the coronal
optical depth to $\tau\approx1$ and a rise in the coronal temperature
to above 80 keV - Maccarone \& Coppi 2003b).  It is commonly held lore
that the cutoff energies of neutron star spectra in hard states are
typically lower than those for black hole spectra, and typically about
30 keV or less (Zdziarski et al. 1998).  In fact, for many
systems, the true low/hard state occurs and the spectrum takes a form
quite similar to the low/hard states of black holes (Barret et
al. 2000).  While clearly substantial hardening to the spectrum has
occured in 4U 1705-44, the spectrum has entered only an intermediate
state, and after this point, the luminosity begins rising and the
spectrum begins softening again.  Based on this data, we can say only
that the state transition luminosity should be less than about
$7\times10^{36}$ ergs/sec.  Furthermore, if the accreted material is
helium rich, the distance estimate of Haberl \& Titarchuk (1995) drops
to 7.0 kpc, and the luminosity drops to $6.3\times10^{36}$ ergs/sec,
which corresponds to 3.3\% of the Eddington luminosity for a 1.4
$M_\odot$ neutron star.  Because a full state transition is not seen,
we do not include this source in either the table or any of the
calculations based on the table.

{\it 4U 0614+09.}  There does not exist a particularly well-sampled
state transition for this source, but evidence seems to suggest that
the hysteresis effects for it are rather mild.  Barret \& Grindlay
(1995) found that the source was in a low/hard type state during two
EXOSAT observations where the flux was $1.1 \& 1.2 \times 10^{-9}$
ergs/sec/cm$^2$ from 1-20 keV, in an intermediate state when the flux
was $1.5\times10^{-9}$ ergs/sec/cm$^2$ and in a much softer state at
$3.5\times10^{-9}$ ergs/sec/cm$^2$.  We thus take the transition flux
to be $1.3\pm0.2\times 10^{-9}$ ergs/sec/cm$^2$.  The distance has
only an upper limit of 3 kpc from sub-Eddington Type I bursts.  When
correcting to the bolometric luminosity, we assume that the spectrum
has a cutoff energy of about 60 keV, in agreement with typical results
from other neutron star systems, and use the measured power law photon
index of $\Gamma=1.9$.  The bolometric luminosity is then
$2\times10^{36} (d/3kpc)^2$ ergs/sec, or less than about 1\% of the
Eddington luminosity.

{\it 4U 1820-30.}
This source has a known distance due to its association with the
globular cluster NGC 6624.  Its transition luminosity cannot be
measured, however, because the sampling of the pointed RXTE
observations of it is insufficient.  It did appear to show a state
transition in 1997, but the timespan between the last high/soft state
observation and the first low/hard state observation was about 20
days, sufficient time for a rather large flux change.  Given that the
count rate did not continue decreasing after the source entered the
low/hard state, one cannot even be certain that the transition flux is
between the two observed fluxes.  This source is a good candidate for
future monitoring campaigns, as its distance is well known and it is
known to exhibit state transitions.

{\it SLX 1732-304.} This system represents a similar case to 4U
1820-30.  It is a globular cluster source, located in Ter 1 (Parmar et
al. 1989), but has been poorly sampled by {\it RXTE}; only four
observations have been made, all in a low flux state, and showing
similar X-ray spectra (Molkov et al. 2001).  Observations from {\it
Granat} did show a spectral state transition, but there were only two
observations made, a month apart in time, and a factor of four apart
in flux (Pavlinsky et al. 2001).  Using the typically quoted 5.2 kpc
distance to the globular cluster (Ortolani et al. 1999), and the 3-20
keV fluxes of $6.95\times10^{-10}$ and $1.64\times10^{-10}$
erg/sec/cm$^{2}$, the luminosities are found to be
2.25$\times10^{36}$ ergs/sec and 5.0$\times10^{35}$ ergs/sec
respectively, indicating a state transition between 1\% and 0.25\% of
the Eddington rate.  However, no bolometric corrections have been made
to these values, and given the rather large neutral hydrogen columns
to the sources, the corrections are rather uncertain, but should be of
order a factor of 2-4.  Given that the spectral data and the sampling
in time are insufficient to make accurate measurements, we choose not
to include these transition luminosities in the analysis.

{\it KS 1731-260.}
This source has been assumed to lie at the Galactic Center.  Neutron
star model atmospheres for this source in quiescence are consistent
with a distance of about 8 kpc, but the best fitting value is half
that distance, assuming a neutron star radius of 12 km (Rutledge et
al. 2000).  We adopt a distance of $4\pm2$ kpc for this source based
on the neutron star model atmospheres.  Its transition appears to have
occured at a luminosity of about $3.3\times10^{36}$ ergs/sec, based on
the PCA/HEXTE measurements taken on May 25, 1999.  However, with the
distance uncertainty of a factor of $\sim2$, we do not include this
source in calculations.

{\it 4U 1636-53.}
State transitions were first seen in 4U 1636-53 with EXOSAT (e.g. Prins
\& van der Klis 1987), but only the hard-to-soft transition was seen
in these data.  A more recent campaign (RXTE proposal 60032) shows
some evidence that a soft-to-hard state transition probably occurred
between 17 September 2001 and 30 September 2001, as the count rate is
dropping and the spectrum is hardening in the well sampled region of
the lightcurve leading up to 17 September 2001.  The spectrum on 17
September 2001 is well represented by a thermal Comptonization model
with a temperature of 3.7 keV and an optical depth of 4.7, so the
source is clearly still in an intermediate state.  Unfortunately, no
observations were taken between 17 and 30 September, and the count
rate had again started rising by September 30, so the transition
cannot even be said to have occurred in a flux range bracketed by the
values of the fluxes on these dates.  We can thus estimate only an
upper limit on the transition flux for this source of the
$1.5\times10^{-9}$ ergs/sec/cm$^{2}$ seen on 17 September 2001.  Given
the 5.5 kpc distance to the source, estimated from radius expansion
Type I bursts (van Paradijs et al. 1986; van Paradijs \& White 1995),
we estimate that the state transition luminosity occurs at no more
than $4.8\times10^{36}$ ergs/sec, implying that the Eddington fraction
of the state transition is less than 2.5\%.

{\it Cir X-1} Circinus X-1 is an extremely unusual system.  It has
been tentatively classified as an atoll source (Oosterbroek et
al. 1995), but shows much larger luminosity variations than most
systems in its class.  Furthermore, its mass accretion rate seems to
vary due to the eccentricity of its orbit, and not due to disk or mass
transfer instability effects that affect most other neutron star
systems.

{\it GX 3+1} This system has not been the subject of a regular, finely
sampled monitoring campaign by {\it RXTE}.  The only regular
monitoring, in Proposal 60022, was roughly monthly.

{\it 4U 1735-44} This system is thought to be at a distance of 9.2 kpc
(van Paradijs \& White 1995) from Type I X-ray bursts, but again there
are no well sampled state transitions observed.  Furthermore, the ASM
lightcurves show that the luminosity, assuming this distance, seems
not to drop below about 15\% of the Eddington rate, so in fact, it
seems likely that no state transition has occurred during the {\it
RXTE} mission.

{\it The persistently bright Galactic Center sources.} Four atoll
sources near the Galactic Center, GX 3+1. GX 9+1, GX 9+9, and GX 13+1
have generally been found only in the banana branches after rather
detailed studies (see Homan et al. 1998 and references within).  These
sources hence do not undergo state transitions, and some have even
sometimes been classified as Z sources rather than as atoll sources
(see e.g. Schulz et al. 1989).  These sources are thus rather similar
to 4U 1735-444.

{\it XTE J 2123-058.}  This source is a new transient first detected
with {\it RXTE} in June of 1998.  It was observed with the pointed
instruments on {\it RXTE} five times, including on either side of a
state transition.  Unfortunately, the flux dropped by a factor of 15
between the lowest soft state and highest hard state points (Tomsick
et al. 1999).  Given the flux measurements of $1.1\times10^{-9}$ and
$7.3\times 10^{-10}$ respectively, along with the distance estimate of
$8\pm3$ kpc (Zurita et al. 2000), we find that the transition occurred
between 14\% and 0.2\% of the Eddington rate (accounting for both the
distance and flux uncertainties).  This range is not particularly
useful, so we exclude this source.

{\it XTE J1806-276} This source has no reliable distance estimate;
there is a only a tentative association with an X-ray burster
(Marshall et al. 1998), and there is only a ``likely'' optical
counterpart (Hynes et al. 1998).

{\it 4U 1915-05} This system was monitored by {\it RXTE} with quite
good sampling in May of 1996, but the well sampled part of the light
curve showed only the upper and lower banana states (Boirin et
al. 2000).  The other data for this source is sampled typically
monthly.  Thus no state transition has been observed in a well sampled
data set.

{\it 4U 1724-30} This system is in Terzan 2, a globular cluster about
9.5 kpc away.  During the {\it RXTE} mission it reached a rather large
luminosity (about 10\% of the Eddington rate), where it was observed
rather frequently (P20170, PI:Jung), but did not show a state
transition; the spectrum always remained hard, even at the peak (as
determined from the long-timescale lightcurve in Emelyanov et
al.(2002), being well fit by a thermal Comptonization model with an
optical depth of about 0.5 and a temperature of about 45 keV.

{\it 4U 1746-37} This is another globular cluster system lacking a
well-sampled state transition luminosity measurement.  There are two
well sampled {\it RXTE} campaigns on this source, P30701 (PI:van
Paradijs) and P60044 (PI:Smale), but P30701 has been shown to be in
the island state all the time (Jonker et al. 2000), and an inspection
of the data from P60044 has shown that the same is true here.  An
additional set of observations, P10112, that was less well sampled
showed the source only in an island state, and occurred well before
the other two campaigns (Jonker et al. 2000).

\subsubsection{Black hole sources with insufficient optical data}

{\it GX 339-4.} 
GX 339-4 is in some ways the best candidate for studying state
transitions - it is one of the few sources which has been well studied
in the X-rays in all the canonical spectral states.  However, its
optical counterpart is very faint, and its lowest luminosity
observations are still dominated by the accretion flow in the optical
and infrared.  Thus no mass measurement has been possible and the
distance measurements to this source are based primarily on the
absorbing medium.  Recent optical studies (Hynes et al. 2003) of the
source in outburst have allowed the measurement of a {\it mass
function}, i.e. a lower limit on the mass, which are sufficient to
prove that the primary is a black hole under the standard assumptions
about the neutron star equation of state, but the mass and distance
estimates remain insufficient for estimating the Eddington luminosity
of the state transitions.

{\it XTE J 1859+226.}
This source lacks an inclination angle estimate, a distance estimate,
and is the subject of a debated period.  The mass function is hence
uncertain and regardless of its certainty, it cannot be converted to a
mass measurement.  

{\it GRS 1009-45.} 
Three problems exist for this source.  The inclination angle is
extremely uncertain, so the conversion from mass function to mass is
likewise uncertain.  The H$\alpha$ emission does not follow the
spectroscopic phase in the conventional manner, suggesting that the
orbital period may be in error.  Finally, the spectral type of the
mass donor is highly uncertain, so there is no distance estimate
(Fillipenko et al. 1998).

\subsubsection{Black hole sources with insufficient X-ray data}

{\it 4U 1543-47.} This source is a dynamically confirmed black hole
candidate, and has shown three strong X-ray outbursts.  Unfortunately,
the first occurred in 1971, when it was monitored once a month over
the entire outburst, the second in 1983, when it was monitored more
frequently, but only during the high/soft state, and the third
occurred in 1992, where the only publicly available data set comes
from BATSE, which is not well-suited to measuring the bolometric
luminosity, and whose data results are available only as 1 week
averages.  It is currently in its fourth outburst, and the current
outburst is likely to provide useful data.

{\it A0620-00.} This source was well observed for two months during its
outburst, but its spectrum only softened throughout the outburst.
This softening is generally consistent with the decrease in
temperature of an accretion disk which is dominating the X-ray
spectrum.  Hence the state transition was not observed in the
well-sampled portion of the light curve and the transition flux cannot
be estimated.

{\it XN Oph 77.} This system was monitored during its outburst by Ariel
V, but the published results present only count rates.  Hence the
state transition cannot be detected and its flux cannot be estimated.

{\it V 404 Cyg.} This source was observed only by {\it Ginga}.  The
data are not public, and the published results (Kitamoto et al. 1989)
are insufficient for determining a time of the state transition.  It
appears that the time resolution of the observations is regardless
insufficient for determining a transition luminosity.

\subsubsection{Black hole sources which do not show the necessary state transitions}

A handful of black hole X-ray transients have not shown the full
phenomenology of spectral states seen in the typical soft X-ray
transients.  As a result, there is no soft-to-hard state transition
for these systems, so the measurement obviously cannot be made.

{\it GRO~J~0422+32 \& XTE~J~1118+480.} 
Both these systems exhibited ``mini-outbursts'' in which the
luminosity never reached a high enough level to trigger a high/soft
state.  

{\it V4641~Sgr.} 
This source was seen to spend a large amount of time in the low/hard
state in the period before a very rapid, very bright outburst on
September 15-16, 1999.  It may have entered the high/soft state on the
way down from the $\sim L_{EDD}$ peak, but the luminosity decay was so
rapid that even the several pointings per day from the {\it RXTE} All
Sky Monitor provide insufficient time resolution to determine whether,
or at what luminosity, a state transition occurred.

{\it GRS~1915+105.}  Since its discovery by {\it Granat}
(Castro-Tirado et al. 1992), GRS 1915+105 has shown a luminosity
consistently sufficient to place it in the high/soft state, the very
high state, or an unstable flaring state.  In fact, its {\it lowest}
luminosity observations are those in the high/soft state.  At times,
it enters into a state commonly referred to as its ``low/hard'' state,
but the spectral index of the power law component in this state is
typically about $\Gamma=2.5$, much softer than any of the other
low/hard states.  It has been suggested that at high fractions of the
Eddington luminosity, flows with high viscosities may see a large
fraction of their energy dissipated in a hot corona (Merloni 2003); in
this picture, in fact, two solutions, a disk-dominated and a corona
dominated one, are possible at every accretion rate above a critical
fraction of the Eddington rate which is a function of the viscosity
parameter $\alpha$.  The properties of these corona are still not well
studied, but the presence of an alternative means of explaining a
power-law dominated spectrum at high accretion rates lends credence to
the suggestion that the GRS 1915+105 ``low/hard'' state is
fundamentally different than that seen in other systems.

\subsection{Bolometric corrections}
Bolometric corrections are applied to the data assuming a spectrum of
$\frac{dN}{dE}\sim$$E^{-1.8}\exp^{-E/200 \rm{keV}}$, and that this
functional form applies from 0.5 keV to 10 MeV, with no power outside
this range.  This model provides a phenomenologically good fit to the
broadband spectra observed from low/hard state objects, and this
correction is essential for correctly estimating the observed
luminosities in the low/hard state (see e.g. Zdziarski et al. 2002).
Changes in the spectral index by 0.1 or changes in the high or low
energy cutoff by factors of two induce only 10\% errors in the
bolometric luminosity, so the errors in the bolometric luminosity
correction are likely to be small compared to the other errors in the
problem.  We note that because the errors in the bolometric
corrections are likely to be rather small, good temporal coverage of
the state transitions becomes more important for estimating the
transition luminosity than good broadband spectral coverage.  Thus,
when multiple data sets exist for measuring the most recent state
transition from a given system, we have chosen the data set with the
best temporal sampling of the source, rather than the one with the
best broadband spectroscopy.

\subsection{Reliability of optical distance measurements}
To make a distance estimate of a black hole or a neutron star from the
properties of its optical counterpart, one needs accurate measurements
of the period and inclination angle of the binary system, as well as
the spectral type of the optical star and the reddening along the line
of sight to the X-ray binary.  The radius of the mass donor is then
assumed to be that at which it exactly fills its Roche lobe (its mass
having been estimated from its spectral type).  The absolute magnitude
is then computed from the spectral shape for the companion star and
the radius estimates, and the distance is estimated from comparing the
observed apparent magnitude with the estimated absolute magnitude.
The models also include a correction for the optical flux contributed
by the accretion flow.  A good discussion of converting optical
measurements into masses and distances can be found in Orosz \& Bailyn
(1997).

The reliability of optical distance measurements has been studied by
Barret et al. (1996).  They find that the largest source of
uncertainty in the mass and distance measurements typically comes from
the uncertainties in the inclination angle.  They found that using the
method above, the uncertainties in the mass are typically $\sim10\%$
for inclination angle errors of about 10 degrees.  They tested this
claim by using an alternative model to estimate the size of the mass
donor - the period density relation from Frank et al.  (1992). They
confirmed that the scatter in the distance measurements was about
15\%, but at least some of this error may be due to the fact that the
mass donors were assumed to be 0.4 $M_\odot$ in order to apply this
technique.

Given improvements in both the quality of the photometric light curves
and the techniques of the ellipsoidal modulation modeling in the last
7 years, the largest uncertainties often come from other
considerations, typically the reddening or spectral type of the mass
donor; typical errors in the inclination angle are now less than 5
degrees (see e.g. Gelino et al. 2001; Orosz et al. 2002).  For the
three sources in our sample where the uncertainties in the optical
distance estimate are greater than 20\%, the distance errors are
largely due to the uncertainties in the spectral type of the companion
star, and not due to errors in the inclination angle.

\subsection{Reliability of burst distance measurements}
Several of the neutron star source distances were measured using Type
I bursts.  It has long been believed that these bursts should be
Eddington limited and hence the brightest bursts observed from a
particular source should be a standard candle.  Recently, this claim
has been tested systematically by examining the luminosity
distribution of Type I bursts in globular cluster X-ray binaries.  In
the globular clusters, of course, the distances can be measured by
main sequence fitting of the optical stars.  It was found that
distances measured from Type I bursts have systematic biases of up to
50\%, depending, for example on whether the accreted material is
hydrogen rich or hydrogen poor, and on whether the truly ``brightest''
radius-expansion burst from a particular system has been observed (see
Kuulkers et al. 2003).  Given that the errors can be as large as 50\%,
but that only a fraction of the sources are likely to have errors this
large, we estimate the typical distance errors due to using the Type I
burst method to be 30\%.

\section{Discussion} 

\subsection{Variations of the transition luminosities}
We have performed averages of the state transition luminosities in
Eddington units weighted by the inverses of their standard deviations,
and also estimate their standard deviation.  We perform the
statistical analyses for three cases - one where Aql X-1 is eliminated
(because, as a neutron star system, its state transitions may depend
differently on luminosity than those of the dynamically confirmed
black hole candidates), one where Aql X-1 is included and the 2.5 kpc
distance of Chevalier et al. (1999) is used, and one where Aql X-1 is
included and the 4.0 kpc distance of Rutledge et al. (1999) is used.

When only the black hole candidates are considered, the weighted mean
state transition luminosity is $1.9\pm0.2$\%, and the excess deviation
(i.e. the square root of the weighted variance minus the weighted
errors) is about 0.57\%, indicating that there are $\approx$ 30\%
variations in the state transition luminosity.  For the purposes of
computing the excess deviation, the error in the state transition flux
has been assumed to be 20\%.  With the neutron stars included, the
mean state transition luminosity drops to $2.2\pm0.2$\% (regardless of
which distance estimate is chosen for Aql X-1), with a fraction excess
variance of 42\% of the mean if the 5.2 kpc distance is used and of
48\% of the mean if the 2.5 kpc distance is used.  Thus there is some
suggestive evidence that the state transitions vary based on the
sample as a whole, although if the 20\% estimate of the systematic
error on the state transition flux is a severe underestimate, then the
sample could conceivably be consistent with measurement errors.
Furthermore, the errors in estimating the variance are likely to be
quite large since there are only 6 black hole candidates in the sample
and the errors on the measurements are quite large themselves.  If the
sources are distributed uniformly in the cosine of the inclination
angle $i$ and the variations in the transition luminosity are entirely
due to bulk relativistic motions of the X-ray emitting region, then
one would expect 30-35\% variations in the state transition luminosity
for a value of $\beta$ between 0.18 and 0.22 (where $\beta$ is
defined as the jet velocity divided by the speed of light).

An interesting question would be whether there is a systematic
difference between the state transition luminosities of black hole and
neutron star accretors.  In fact, the weighted mean state transition
luminosity for the neutron stars is larger than that of the black
holes by a rather substantial margin.  However, we note that there are
only 3 neutron stars with suitable measurements for inclusion in the
sample, so random variations are likely to be quite important.
Furthermore, there are two more neutron stars with upper limits for
the state transition luminosity that are substantially below those of
the neutron stars included in the sample.  At this time the data are
insufficient for making a strong statement on this question.

Stronger evidence of real variations may be found by looking at
individual sources.  In particular, the source GRO~J~1655-40 has a
state transition at a luminosity $4\sigma$ below the mean state
transition luminosity.  This system has one of the lowest observed
state transition luminosities in the sample, and has the smallest
observational errors, largely due to strong constraints on the
distance given by the combination of the relativistic jet kinematics
and the optical distance measurement, and the strong constraints on
the orbital inclination angle given by the combination of strong
ellipsoidal modulations and the lack of an eclipse (Greene et
al. 2001).  Furthermore, it has a well constrained jet inclination
angle of 85 degrees (Hjellming \& Rupen 1995).  While the errors
listed in Hjellming \& Rupen (1995) are likely to be underestimated,
it has been shown that the observed jet proper motions are
incompatible with the minimum distance constraints from optical
measurements of 3.1 kpc if the jet's inclination angle is less than
about 82 degrees (Fender 2003).

Because the inner accretion disk of GRO~J~1655-40 is so close to being
edge-on, if beaming effects are important, GRO~J~1655-40 is likely to
have its hard X-ray luminosity de-beamed, and hence to have a lower
transition luminosity than the mean.  We define the approaching beam's
Doppler factor $\delta_{app}$ to be $[\Gamma(1-\beta cos
\theta_i)]^{-1}$ and the receeding beam's Doppler factor
$\delta_{rec}$ to be $[\Gamma(1+\beta cos \theta_i)]^{-1}$.  We
average the beaming factor $\delta_{app}^{2.8}+\delta_{rec}^{2.8}$
(from assuming a $\Gamma=1.8$ spectral index and a continuous flow of
material into the emission region and applying the appropriate formula
from the review of Mirabel \& Rodriguez 1999), over angles to
determine the ratio between the intrinsic rest frame and mean observed
fluxes.  We find that for the most likely value of the GRO~J1655-40
state transition luminosity of 0.95\% $L_{EDD}$, $\beta=0.62$.  For
the lowest reasonable value, where all the errors are set to the
$1\sigma$ value in the direction where this acts most strongly to
reduce the inferred transition luminosity of GRO~J~1655-40, we find
that $\beta=0.71$, making this value the upper limit on the velocity
of the emission region.  Taking instead the maxmimum value for the
distance and the minimum value for the mass, and assuming the state
transition flux was {\it overestimated} by about 20\%, we find that
the state transition luminosity was still no more than about 1.4\% of
the Eddington luminosity.  If the variations in state transition
luminosity are then solely due to relativistic beaming, we would
expect the outflow velocity to be at least $\beta=0.51$.

We note that there has been a recent challenge to the standard
distance estimate for GRO~J~1655-40 (Mirabel et al. 2002).  These
authors have argued most strongly that the optical distance
constraints may not be valid, as one cannot be sure that a star in an
interacting binary has the same luminosity as a single star of the
same spectral type.  They have furthermore suggested that perhaps the
relativistic jet and the outer accretion disk in this system are
aligned with one another, which would place the source at a distance
of about 0.9 kpc assuming that the proper motions of the jet have been
measured accurately.  Then, however, the state transition for
GRO~J~1655-40 would occur at about $7\times10^{-4} L_{EDD}$, and if
the differences were solely due to beaming effects, the X-ray emission
region's Lorentz factor in the low/hard state would have to be about
$\Gamma>3.0$, large enough to violate severely the beaming constraints
based on the radio luminosity to X-ray luminosity correlation of
Gallo et al. (2002).  Additionally, a higher beaming factor in the
low/hard state would be implied than in the ``superluminal'' jets if
the source were at such a small distance, in seeming contradiction to
the observation that the jets of the low/hard state systems tend to be
slower than those seen in the high luminosity flaring states (e.g.
Stirling et al. 2001).  Finally, it seems unlikely that a single
source would be separated from the sample as a whole by a factor of
$\sim20$ in state transition luminosity while the rest of the sample
is clustered within a factor of three of the mean, despite rather
large measurement errors which should lead to larger dispersion in
their transition luminosity values.

Finally, we point out that while the X/$\gamma$-ray luminosities of
most sources with superluminal non-steady jets are very close to the
Eddington limit at the time of the jet event, GRO~J~1655-40 was
emitting at only about 70\% of $L_{EDD}$ when its jet was ejected
(Sobczak et al. 2000).  It is not clear, however, whether this
represents evidence for relativistic motions in the very high state.
The X/$\gamma$-ray spectrum of sources in this state have a
substantial contribution from a geometrically thin, optically thick
component, so simple geometric projection effects can lead to changes
in flux received as a function of the observer's inclination angle.

\subsection{Implications for theoretical models}
The present observational constraints cannot distinguish among
existing theoretical models.  Three broad classes of geometries exist
for explaining the emission.  One class, that of Comptonization in a
geometrically static, optically thin medium (e.g. Shapiro et al. 1976
and other similar solutions such as ADAFs) predicts that there should
be no variations in the state transition luminosities due to geometric
effects - a static optically thin medium emits isotropically.  In this
picture, the state transition luminosity should show no variations.
Our finding that GRO~J~1655-40 shows evidence for such variations
casts some doubt on the validity of this model.  Given that only one
black hole candidate shows such clear evidence for variations in the
luminosity of the state transition, at this point, we cannot rule out
the possibility that this observational result is due to intrinsic
variations in the state transition luminosities rather than due to
relativistic beaming.  Nonetheless, the case of GRO~J~1655-40 seems to
suggest that the most likely cause of the variations are inclination
angle effects on the state transition luminosities.

The two other possibilities, those of synchrotron emission from jets,
and of magnetic flares in a corona above an accretion disk, are also
optically thin solutions.  However, both these models have bulk
relativistic accretion flows, and the resulting relativistic beaming
can have an effect on the observed luminosity.  In the case of a
corona powered by magnetic reconnections, the bulk outflow velocity
has been estimated on the basis of the relative weakness of the
Compton reflection component in the low/hard state (i.e. the
correlation between the spectral index and the reflection fraction
found by Zdziarski et al. 1999).  Given that $R/2\pi \sim 0.3$ in the
low/hard state, Beloborodov (1999) estimated that the corona would
have an outflow velocity of $\beta \sim 0.3$.  This is in relatively
good agreement with the observational results presented here, as the
value of $\beta$ falls between that suggested by the GRO~J~1655-40
data alone and that suggested by the sample as a whole.

The final mechanism proposed for the emission from low/hard state
objects is emission from a relativistic jet.  In the models of Markoff
et al. (2001), the typical beaming factors inferred are
$\Gamma\sim2.0$, which is equivalent to $\beta\approx0.85-0.9$
(S. Markoff, private communication), which is somewhat larger than the
minimum value of $\Gamma$ suggested by the VLBI measurements of
Stirling et al. (2001) for Cygnus X-1 in its low/hard state.  Such
beaming factors are marginally too large to be in agreement with the
observational data on state transition luminosities, and if the
current results hold up as a larger sample of sources is examined,
inclination angles of the jets of more systems are measured and
smaller error bars are placed on the masses and distances of the
sources, then this result will present fairly strong evidence against
the X-ray jet model in its current form.  Such models, could, however,
be modified (for example by adding the effects of Compton drag, which
should slow the jet).  At the present, the discrepancy is not large
enough to allow for a strong conclusion.

There are no strong constraints on the inclination angles of the jets,
and hence on the inner disks of systems other than GRO~J~1655-40.  The
inclination angles of the binary planes of these systems are generally
constrained within about 20 degrees, but the inner disks are subject
to warping due to, among other effects, the Bardeen-Petterson effect,
and the timescale for the black hole spin to realign itself with the
binary plane is often longer than the lifetime of the binary system
(Maccarone 2002).  It would thus be of great use to measure the
inclination angles of jets from more of these systems, but such
measurements are typically only possible during flaring events where
proper motions can be observed.  Such measurements are, of course, the
key to determining whether the suggestive results from GRO J 1655-40
are truly an indication of an inclination angle effect on state
transition luminosities (and hence beaming), or just a coincidence.

If one of the established mechanisms for producing the spectral state
transitions can be successfully adopted, then the value of $\alpha$
might be measured using the state transition luminosities.  In the
case of advection dominated accretion flows, the state transition
luminosity occurs at a luminosity of 1.3$\alpha^2 L_{EDD}$ (Narayan \&
Yi 1995), so the observations that the state transitions occur at
0.022 $L_{EDD}$ would indicate that $\alpha=0.13$.  The magnetic
corona model of Merloni \& Fabian (2002), on the other hand, suggests
that the state transitions should depend very weakly on $\alpha$, but
should always be at luminosities slightly higher than 2\%.  Other
models do not predict such a clear correlation between $\alpha$ and
the state transition luminosity.

\subsection{A new distance indicator?}
We have found that the state transition luminosity for X-ray binaries,
and particular for the black hole systems is constant to within about
40\%.  This in turn means that given perfect flux and mass
measurements, the state transition flux could be used as a distance
indicator within about 20\% accuracy.  Recent work on X-ray binaries
in bright states has made it possible to estimate their masses in
outbursting states when the stellar spectrum cannot be estimated
accurately (Steeghs \& Casares 2002; Hynes et al. 2003).  In
particular, GX 339-4, whose distance has not been reliably measured,
but is generally assumed to be about 4 kpc on the basis of its
kinematics and its hydrogen column measurements (Zdziarski et
al. 1998), has had its mass function measured by this method; the mass
is found to be at least 5.8 $M_\odot$.  Its state transition flux,
$F_{trans}$ has been measured by Nowak, Wilms \& Dove (2002), and is
found to be $2.7\times10^{-9}$ ergs/sec/cm$^2$.  If the state transition
luminosity is 2.2\% of the Eddington limit, then it should be at least
$1.8\times10^{37}$ ergs/sec for a black hole with $M>5.8M_\odot$.
This distance should be $d>\sqrt(L/4\pi$$F_{trans}$), or greater than
7.6 kpc.  Given a likely 40\% systematic error on the state transition
luminosity in Eddington units, a 20\% error on the flux measurement,
and a 9\% error on the mass function measurement, the errors on the
distance are likely to be about 22\%, so we find that the distance to
GX 339-4 is at least $7.6\pm1.8$ kpc by this method.  Applying a
similar argument to the recent very strong flare of GX 339-4, where
the source reached a flux of 1.4$\times10^{-7}$ ergs/sec/cm$^2$ (as estimated
from the {\it RXTE} All Sky Monitor flux on 16 July 2002), and
assuming that it reached the Eddington luminosity for a $>$ 5.8
$M_\odot$ black hole, we find a distance that must be at least 7.1
kpc.  A distance of more than 7 kpc is perhaps not surprising
considering that the source lies along a line of sight that passes
near to the Galactic Center.

\subsection{Prospects for future improvements}
Finally, we note that these results underscore an additional important
point - the need for more accurate distance measurements of X-ray
binaries.  The largest uncertainties in the problem are almost always
those in the distance.  As there will always be some systematic
uncertainty in determining the appropriate spectral type for a Roche
lobe overflowing star, an additional method for measuring distances
would be of great benefit.  Parallax distance estimates from planned
interferometers such as SIM, and from VLBI measurements hold great
potential for improving our measurement accuracy of the state
transition luminosities.  Many X-ray binaries approach high enough
radio flux levels in their low/hard states for VLBI measurements to be
made (see e.g. Bradshaw et al. 1999), and hopefully a large fraction
will be observed by SIM.  Great improvements in understanding the
state transition luminosities of neutron star systems could result
from improved temporal sampling of the outburst cycles of globular
cluster X-ray sources since the distance estimates for most globular
clusters are substantially more accurate than those for the individual
stellar companions in X-ray binaries and are free of the systematic
uncertainties that cause problems for Type I burst distance
measurements.  A more sensitive all sky monitor in X-rays would be a
key for such work, since the ASM on {\it RXTE} is not sensitive enough
to easily detect outbursts from the faint sources in the more distant
globular clusters.

\begin{table*}
\begin{tabular}{lccccccccc}

Source& $M$ (in $M_\odot)$& $d$ (in kpc)& $L_{trans}$ (in ergs/s)& $L_t/L_E$& $i$ & $ \sigma_L/L$\cr 
\hline 
Nova Mus 91& 7.0$\pm$0.6&5.1$\pm0.7$ & 3.0$\times10^{37}$ & 0.031& & 0.35\cr 
XTE J 1550-564&10.0$\pm1.5$& 5.9$\pm2.8$ & 4.8$\times10^{37}$ & 0.034 & &0.98\cr 
GS 2000+251& 8.5$\pm1.5$& 2.0$\pm1.0$ & 7.4$\times10^{36}$ &0.0069& & 1.0\cr 
Cyg X-1 & 13.0$\pm3.0$ & 2.5$\pm0.5$ & 4.7$\times10^{37}$& 0.028 & & 0.50\cr 
GRO~J~1655-40& 6.3$\pm0.5$ & 3.3$\pm0.2$ & 9.5$\times10^{36}$& 0.0095 &85 & .25\cr 
LMC X-3 & 9.5$\pm2.0$ & 51$\pm1.0$ & 2.0$\times10^{37}$ &0.014 & & 0.29\cr 
Aql X-1(C) & 1.4$\pm0.1$ & 2.5$\pm0.5$ & 7.0$\times10^{35}$ &0.004 & &0.45\cr 
Aql X-1(R) & 1.4$\pm0.1$ & 5.2$\pm1.2$ & 1.8$\times10^{36}$ &0.019 & &0.51\cr
4U 1608-52 & 1.4$\pm0.1$ & 3.6$\pm1.8$ & 8.1$\times10^{36}$ &0.042 & &0.64\cr
4U 1728-34 & 1.4$\pm0.1$ & 4.3$\pm2.2$ & 9.6$\times10^{36}$ &0.050 & &0.64\cr
\end{tabular}
\caption{The table of the source parameters - compact object mass in
$M_\odot$, binary system distance in kpc, transition luminosity in
ergs/sec, assuming isotropic emission and the distance in the previous
column, ratio of transiton luminosity to the Eddington luminosity,
inclination angle, and fractional error on the state transition
luminosity, assuming standard error propagation.  The Aql X-1(C)
refers to Aql X-1 using the distance estimate of Chevalier et
al. (1999), Aql X-1(R) refers to Aql X-1 using the distance estimate
of Rutledge et al. (1999).}
\end{table*}

\begin{acknowledgements}
We wish to thank Rob Fender, Elena Gallo, Sera Markoff and Andrea
Merloni for useful discussions.  This research has made use of data
obtained through the High Energy Astrophysics Science Archive Research
Center Online Service, provided by the NASA/Goddard Space Flight
Center.
\end{acknowledgements}

\end{document}